\documentclass[prc,twocolumn,superscriptaddress]{revtex4}
\usepackage{amsmath,amssymb,epsfig,bm}
\newcommand{\be}{\begin{equation}}
\newcommand{\ee}{\end{equation}}
\newcommand{\bea}{\begin{eqnarray}}
\newcommand{\eea}{\end{eqnarray}}

\begin{document}
\title{Equilibrium sequences of non-rotating and rapidly
rotating crystalline color-superconducting hybrid stars
}
\author{Nicola~D.~Ippolito} 
\affiliation{
 Dipartimento di Fisica, Universit\`a di Bari 
and  I.N.F.N., Sezione di Bari, I-70126 Bari, Italia}

\author{Marco~Ruggieri}
\affiliation{
 Dipartimento di Fisica, Universit\`a di Bari 
and  I.N.F.N., Sezione di Bari, I-70126 Bari, Italia}

\author{Dirk~H.~Rischke} 
\affiliation{ Institut f\"{u}r Theoretische Physik,
J.~W.~Goethe-Universit\"{a}t, D-60438 Frankfurt am Main, Germany}
\affiliation{Frankfurt Institute for Advanced Studies,  
J. W. Goethe-Universit\"at, D-60438 Frankfurt am Main, Germany}

\author{Armen~Sedrakian}
\altaffiliation[Also at ]{Institute for Theoretical Physics, T\"ubingen University.}
\affiliation{ Institut f\"{u}r Theoretische Physik,
J.~W.~Goethe-Universit\"{a}t, D-60438 Frankfurt am Main, Germany}

\author{ Fridolin Weber}
\affiliation{ Department of Physics, San Diego State University,
San Diego, CA 92182-1233, USA}

\date{\today}

\begin{abstract}
The three-flavor crystalline color-superconducting (CCS) phase 
of quantum chromodynamics (QCD) is a candidate phase for the ground 
state of cold matter at moderate densities above the 
density of the deconfinement phase transition. Apart from being a superfluid, 
the CCS phase has properties of a solid, such as a lattice structure and
a shear modulus, and hence the ability to sustain multipolar 
deformations in gravitational equilibrium. We construct equilibrium 
configurations of hybrid stars composed of nuclear matter at low, and 
CCS quark matter at high, densities. Phase equilibrium between these
phases is possible only for rather stiff equations of state of nuclear 
matter and large couplings in the effective Nambu--Jona-Lasinio Lagrangian 
describing the CCS state. We identify a new  branch of stable CCS hybrid 
stars within a broad range of central densities which, depending on the
details of the equations of state,  either bifurcate 
from the nuclear sequence of stars when the central density exceeds 
that of the deconfinement phase transition or form 
a new family of configurations separated from the purely nuclear sequence
by an instability region. The maximum masses of our non-rotating
hybrid configurations are consistent with the presently 
available astronomical bounds. The sequences of hybrid configurations 
that rotate near the mass-shedding limit are found to be more compact 
and thus support substantially larger spins than their same mass nuclear 
counterparts.

\end{abstract}


\maketitle

\section{Introduction}

Matter in the interiors of neutron stars is compressed by gravity to
densities which are by factors 5 to 10 larger than the density of an
ordinary nucleus. At such densities baryons are likely to lose
their identity and dissolve into their constituents - deconfined
quarks. It has been suggested long ago that quark matter may exist 
in the interiors of compact objects~\cite{DECONFINMENT}. 
Within the deconfined phase
unbound quarks form  Fermi seas and attractive interactions
bind them into Cooper pairs thus giving rise to quark
superconductivity~\cite{CSC,Casalbuoni:2003wh}. 
Along with the deconfinement phase
transition the dense matter undergoes a second transition related to
the  restoration of  chiral symmetry (the dynamical symmetry of the
strong interaction). This symmetry may remain broken in some 
color-superconducting phases, such as the color-flavor-locked phase.

If compact (hybrid) stars featuring quark cores surrounded 
by a nuclear mantle exist in nature, they could provide a
unique window on the properties of QCD at high baryon densities.
Even if the relevant densities and sufficiently 
low temperatures could be
reached in a laboratory in the future, the conditions prevailing in
compact stars are different from those produced in accelerators:
the matter is long-lived, charge neutral and in
$\beta$-equilibrium with respect to weak interactions.

A central question of the theory of compact stars, which we will
address in this work, is whether the equation of state of matter at
high densities admits stable configurations of self-gravitating
objects in General Relativity featuring deconfined quark matter, and
if so, are the gross parameters of these objects, like their mass and
radius, compatible with the known astronomical bounds? The
deconfinement phase transition from baryonic to quark matter leads to a
softening of the equation of state which could lead to an
instability towards a collapse into a black hole. The details of a
stability analysis may depend on the theoretical model of
deconfined, color-superconducting quarks. The
Nambu--Jona-Lasinio (NJL) model is a  non-perturbative low-energy
approximation to  QCD, which is anchored in the low-energy
phenomenology of the hadronic spectrum. While  dynamical symmetry
breaking, by which quarks acquire mass, is incorporated in this model,
it lacks confinement~\cite{Klevansky:1992qe}. Previous studies of
hybrid color-superconducting stars within the three-flavor NJL model
displayed a general instability of hybrid stars towards  collapse
into a black hole~\cite{Buballa:2003et}. 
Stable stars featuring two-flavor superconducting (2SC) phases 
were obtained with typical maximum masses $M\sim 1.7 M_{\odot}$ under 
a favorable choice of constituent quark masses (which are related to 
the value of the baryo-chemical potential at zero 
pressure)~\cite{Shovkovy:2003ps}
or by replacing the hard NJL cut-offs by soft form factors with
parameters fitted to the same set of data~\cite{Grigorian:2003vi,Blaschke:2007ri}.
Heavier objects are obtained if a repulsive vector interaction is 
introduced in the NJL Lagrangian~\cite{Klahn:2006iw}.

More phenomenological studies of hybrid stars based on the MIT bag model 
where carried out using a generic parameterization of the quark matter equation 
of state~\cite{Alford:2004pf}. Non-perturbative QCD corrections to the 
equation of state of the Fermi gas were accounted through the parameter 
$a_4=1-c$, where $c\simeq 0.3$ is a reasonable 
estimate~\cite{Fraga:2001id}. For this value of the correction stable hybrid 
stars containing color-flavor-locked superconducting quark matter exist 
with maximum mass 1.9~$M_{\odot}$.

Because of $\beta$ equilibrium in the light quark sector
the chemical potentials of  $u$ and $d$ quarks obey the constraint
$\mu_d \simeq \mu_u+\mu_e$~\cite{COMMENT1}, 
from which it is evident that the Fermi
surfaces of the light flavor quark flavor are shifted by an amount corresponding
to the chemical potential of electrons.  Furthermore, the Fermi surface
of strange quarks is mismatched with the Fermi surfaces of 
light $u$ and  $d$ quarks because of the large strange quark mass. 
A generic feature of fermionic systems with mismatched Fermi surfaces 
is the existence of phases with broken spatial symmetries. A particular
realization is the Larkin-Ovchinnikov-Fulde-Ferrell (LOFF) phase~\cite{LO,FF}, 
which is theoretically predicted  in different fermionic systems 
ranging from dilute atoms~\cite{LOFF_ATOMS}, nuclear matter~\cite{LOFF_NUCL} 
to dense quark matter~\cite{Alford:2000ze,Casalbuoni:2003wh,Rajagopal:2006ig,
Casalbuoni:2005zp,Ippolito:2007uz,Kiriyama:2006ui,Mannarelli:2006fy,Fukushima:2007bj,he:2007}.

The rotational and translation
symmetries are broken by the superconducting LOFF phase, since the
condensate carries a non-zero momentum~\cite{LO,FF}.
The study of crystalline
color superconductivity (CCS) in the context of QCD reveals that this phase
could be the true ground state of deconfined quark matter at intermediate
densities~\cite{Alford:2000ze,Casalbuoni:2003wh,Rajagopal:2006ig,
Casalbuoni:2005zp,Ippolito:2007uz,Kiriyama:2006ui,Mannarelli:2006fy}.
Later on we shall describe a particular realization of the
three-flavor LOFF phase, which is used in our search for stable
sequences of hybrid stars~\cite{Casalbuoni:2005zp,Ippolito:2007uz}.

The nuclear equation of state, as is well known, can be constructed starting
from a number of different principles \cite{WEBER_BOOK,Sedrakian:2006mq}.
We tested a large number of equations of state to construct 
hybrid star configurations. These  
belong to the classes of (i) non-relativistic
variational and Bruckner-Hartree-Fock theories which use as an input
a non-relativistic potential fitted to the elastic nucleon-nucleon
scattering data; (ii) relativistic mean-field models which are fitted to
the bulk properties of nuclear matter, and
(iii) relativistic models which include correlations
at the level of the covariant scattering amplitude 
(Dirac-Bruckner-Hartree-Fock theories). For present
purposes, it is sufficient to characterize these equations
of states by their stiffness; as we shall see, only the stiffest
equations of state are admissible for phase
equilibrium between nuclear and quark matter.

The paper is organized as follows. In Sec.~\ref{sec:phases} we discuss
the input equations of state with the emphasis on the CCS
 phase of quark matter. Section~\ref{sec:results}
contains our results for the sequences of non-rotating and rapidly
rotating hybrid stars. Our conclusions are collected 
in Sec.~\ref{sec:conclusion}.

\section{Phase equilibrium}
\label{sec:phases}

\subsection{The three-flavor crystalline phase of QCD}

\label{subsec:A}
The LOFF phase arises naturally in three-flavor QCD
as a generic feature of a fermionic systems with mismatched Fermi surfaces
~\cite{Alford:2000ze,Casalbuoni:2003wh,Rajagopal:2006ig,
Casalbuoni:2005zp,Ippolito:2007uz,Kiriyama:2006ui,Mannarelli:2006fy,Fukushima:2007bj,he:2007}. The simplest possible realization
of the LOFF phase invokes a single plane-wave modulation of the order
parameter, related to a single value of the center-of-mass momentum
of the Cooper pairs. More complex Ans\"atze for the order parameter
maximize the free energy gain of the superconducting phase for
periodic three-dimensional lattice structures (e.~g.~face-centered-cubic 
lattice in Ref.~\cite{LO} or body-centered-cubic lattice in 
Ref.~\cite{Rajagopal:2006ig}). The studies of complex structures
can be carried out in the Ginzburg-Landau regime, where the order parameter
is small, as is the case near the critical temperature of 
the phase transition or the critical mismatch for unpairing.  
Independent of the form of the lattice structure, the resulting 
CCS state breaks the translational and rotational symmetries since  Cooper
pairs always carry non-zero center-of-mass momentum with respect to some fixed
reference frame. Some of the characteristic features of the CCS state such as a
non-zero shear modulus and gapless excitations imply that it
could play a distinctive role in the phenomenology of compact stars
~\cite{Alford:2000ze,Casalbuoni:2005zp,Rajagopal:2006ig,
Anglani:2006br}.

In the three-flavor case 
the quark condensate is written as a superposition of plane waves
\begin{equation}
\langle \psi_{\alpha i} C \gamma_5 \psi_{\beta j}\rangle \propto
\sum_{I=1} \epsilon_{\alpha\beta I}\epsilon_{ijI} \Delta_I
\sum_{m=1}^{P_I} \exp\left(2i{\mathbf{q}^m_I} \cdot {\mathbf
r}\right)~.\label{eq:ansatz1}
\end{equation}
Here Greek (Latin) indices correspond to color (flavor); $P_I$ is
the number of plane waves of the particular crystalline structure
considered; $\Delta_1,~\Delta_2,~\Delta_3$ refer to $ds$, $us$, $ud$
pairing respectively, and $2\,{\mathbf q}_m^I$ is the quark pair
momentum. The magnitude of ${\mathbf q}_m^I$ and of the gap
parameters $\Delta_I$ are determined by the minimization of the
thermodynamic potential $\Omega$.

The studies of the crystalline structure of the LOFF phase revealed that
there exists a window of the parameter $M_s^2/\mu$, where $M_s$ is
the in-medium strange quark mass and $\mu$ the mean quark chemical potential,
where two crystalline phases, called CubeX and 2Cube45z,
could compete for the ground state of three-flavor quark matter~\cite{Rajagopal:2006ig}.
For both of the structures
$\Delta_1 = 0$, $\Delta_2 = \Delta_3$, $P_2 = P_3 \equiv P$ and
$|{\mathbf q}_2| = |{\mathbf q}_3|$. In the CubeX phase $P = 4$; for
each pairing channel $I=2,3$ the wave vectors $\{{\mathbf q}\}$
point to the vertices of a square, and the two squares are arranged
in such a way that they point to the vertices of a cube. On the
other hand, in the 2Cube25z one has $P=8$; each wave vector set
$\{{\mathbf q}\}$ forms a cube, and the two cubes are rotated by $45$
degrees around an axis perpendicular to one of the faces of the
cube. The analysis of Ref.~\cite{Rajagopal:2006ig} has been extended
in Ref.~\cite{Ippolito:2007uz} by evaluating self-consistently the
strange quark mass for each value of $\mu$, therefore translating
the window of $M_s^2/\mu$ into a window of $\mu$. The
LOFF phases then exist within the chemical potential range 
$442~\rm{MeV}\leq\mu \leq 515~\rm{MeV}$~\cite{Ippolito:2007uz}.
Thus, having the dependence of the pressure on the 
chemical potential, we are in a position to construct phase equilibrium 
and obtain the pressure as a function of the energy density.

A straightforward normalization of the quark pressure in the NJL model
requires that  the pressure  vanishes at zero density
and temperature~\cite{Buballa:1998pr,Sandin:2007zr}. In the terminology of
the MIT bag model, this is equivalent to subtraction of a 
bag constant from the thermodynamic potential~\cite{Alford:2004pf}. 
Since the value of the bag constant is
related to confinement which is absent in the NJL model, it appears
reasonable to consider changes in its value, and hence in the
normalization of the pressure. We shall consider the simple case of a
constant shift in the asymptotic value of the pressure; alternatives
include the use of form factors for the bag constant \cite{Grigorian:2003vi}
or the Polyakov loop at nonzero density~\cite{Fukushima:2003fw,Ratti:2005jh}.

In Ref.~\cite{Ippolito:2007uz} the calculation of masses and condensates
was carried out for $\eta = G_D/G_S= 0.75$, where $G_D$ and $G_S$ are,
respectively, the diquark and the quark-antiquark coupling constants,
and the proportionality factor $\eta$ is obtained via a Fierz
rearrangement of the interaction term in the Lagrangian. This regime
is usually referred to as ``intermediate coupling". Here we shall adopt
a ``strong coupling" regime with 
$\eta = 1 $~\cite{Ruester:2005jc,Sandin:2007zr}, for only in the latter 
case the matching to the nuclear equations state can be performed without
variations in the bag constant (we will discuss this 
point in more detail in the following subsection). 
Clearly, a more complete analysis will
require a (re)computation of the effect of the strong coupling on
the competition among the various superconducting phases such as
the CFL, 2SC and 2-flavor LOFF phases; the possibility of multiple
transitions between these phases is not considered here.

For completeness we reproduce here the parameters of the NJL model 
of Ref.~\cite{Ippolito:2007uz} used in our study: the values of the 
bare $u$, $d$ and $s$ quark masses are $m_u = m_d = 5.5$ MeV and $m_s = 135.4$ MeV, 
the coupling constant for the 4-fermion term of the NJL Lagrangian
is $G_S = 1.81/\Lambda^2$,  the coupling constant for the t'Hooft term
is $K = 8.80/\Lambda^5$, where the ultraviolet cutoff is         
$\Lambda = 643$ MeV. With these parameters the following observables 
of QCD are reproduced: the pion decay constant $f_\pi = 93$ MeV, 
the pion mass $m_\pi = 135$ MeV, the kaon mass $m_K = 497$ MeV, and the 
$\eta'$ mass $m_{\eta'} = 924$ MeV.

\subsection{Matching the equations of state}

\begin{figure}
\begin{center}
\includegraphics[height=8.0cm,width=\linewidth,angle=0]{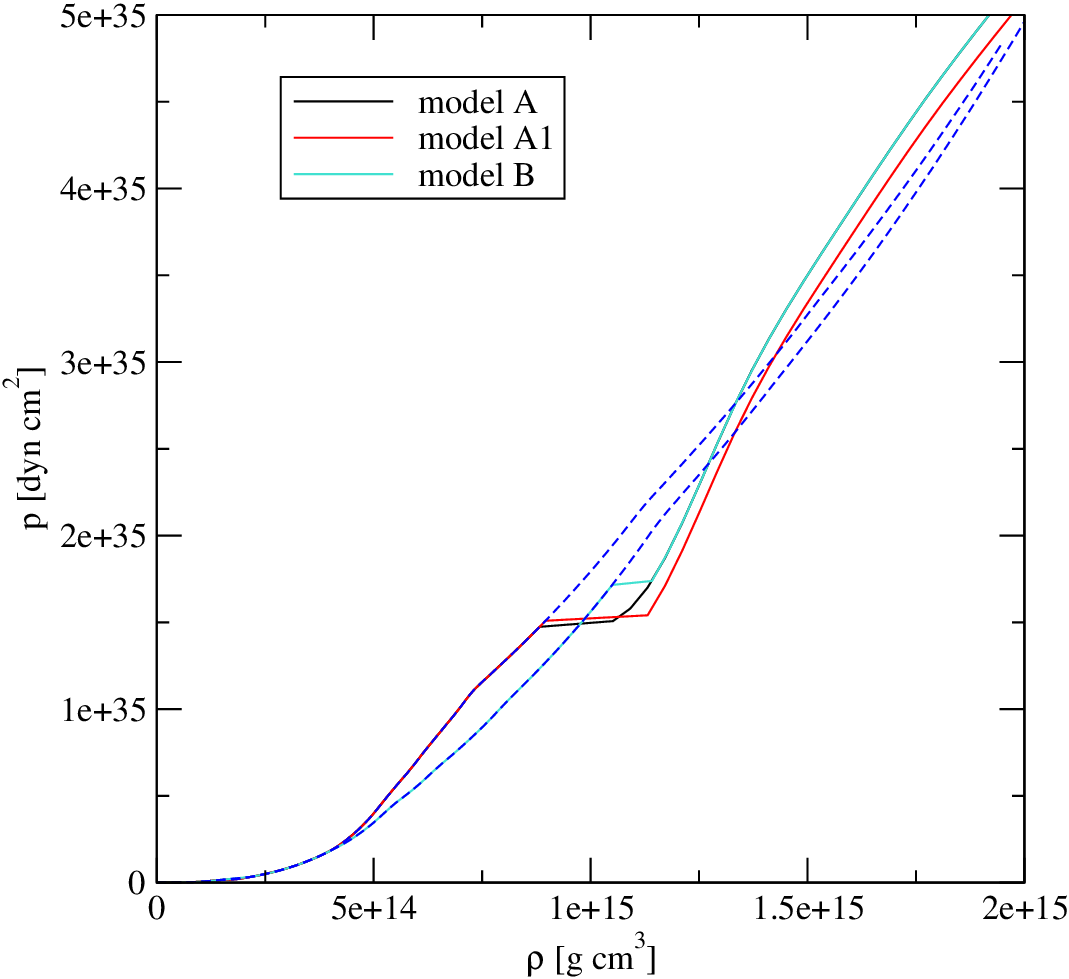}
\end{center}
\caption[]
{Pressure versus density for the models A ({\it heavy, black online}),
A1 ({\it medium-light, red online}), and B ({\it light, blue online}).
The nuclear equations of state are shown by dashed lines
({\it dark blue online}).
For the models A and A1 the  nuclear (low density) equation of
state is the same; for the models A and B the  quark (high density)
equation of state is the same. At  the deconfinement phase
transition there is a jump in the density at constant pressure.
}\label{fig:p_vs_rho}
\end{figure}

Physically, the true nuclear equation state must go over to some 
sort of quark equation of state at some density if deconfinement
takes place in nature. Since we have only models of deconfined 
matter and nuclear matter, this transition is modeled by requiring 
that there exists a baryo-chemical potential at which the pressures 
of these phases are equal. This is equivalent to the condition that
pressure, $P$,  vs. chemical potential, $\mu$, curves for these phases 
cross (matching). If the $P(\mu)$ curves for chosen equations of state 
of nuclear and quark matter do not cross, the models are incompatible
in the sense that they cannot describe the desired transition between 
nuclear and quark matter. The low-density equation of state of nuclear 
matter and the high-density  equation of state of 
CCS matter are matched at an interface via the Maxwell construction.
The phase with largest pressure is the one that is realized at a given
chemical potential. Thus, according to the Maxwell construction of
the deconfinement phase transition, there is a jump in the density
at constant pressure as illustrated in Fig.~\ref{fig:p_vs_rho}.
The high-density regime contains two equations of states for
crystalline color superconductivity which differ by the
normalization of pressure at zero density (or, equivalently, the value
of the bag constant). The model A1 is normalized such that the
pressure vanishes at zero-density. For the models A and B the zero
density pressure is shifted by an amount $\delta p = 10$ MeV/fm$^3$. As
discussed above, this is equivalent to a variation of the bag
constant whose value is related to confinement, is unspecified
in the NJL model, and is uncertain in general. We are aware of the
arbitrariness of the latter procedure, the sole practical purpose of
which is to produce an equation of state which can be matched to a
particular nuclear equation of state. Yet another possibility is to
set $\delta p = 0$, but vary the value of the constituent masses of
the light quarks in the fit of the parameters of the NJL model; for
small values of the light quark masses the matching between  quark and 
nuclear equations of state is
facilitated~\cite{Buballa:2003et}. Furthermore, we need to
adopt a large value for the ratio $\eta$,  e.~g. $\eta=1$, 
since for $\eta= 0.75$ -- the conventional value of this ratio --
no matching occurs between the quark and the 
nuclear equations of state.

A set of nuclear equations of states  were tested for matching
with the models above; it included about  dozen equations of state, 
listed in Refs.~\cite{WEBER_BOOK,Sedrakian:2006mq}. 
Only two equations of state based
on the Dirac-Bruckner-Hartree-Fock approach~\cite{WEBER_BOOK} are suitable 
to match with the quark equations of state presented above.
These are shown in Fig.~\ref{fig:p_vs_rho}.
The selected equations of state are the two most hard equations of state
in the collection. It should be noted that the matching with softer
equations of state can be enforced by varying $\delta p$ by larger
amounts than quoted above. For our purposes, however, the three equations
of state that follow upon matching are sufficient. The values of the
chemical potential (in MeV) and pressure (in MeV/fm$^3$) at the quark-nuclear
matter interface are ($\mu = 1234.2$; $p=96$) for the model A1, (1230.0; 95) 
for  the model A and (1234.8; 108) for the model B.

\section{Results}
\label{sec:results}

\subsection{Non-rotating configurations}
\label{sec:NR-confug}

In this subsection we consider  equilibrium and stability of cold
hybrid stars with CCS cores; rotating
configurations are discussed in the next subsection. Each equation
of state defines a sequence of equilibrium, non-rotating stellar
configurations in General Relativity, which can be parameterized in
terms of the central density $\rho_c$ of the configuration. It is
assumed that the configurations are cold, so that the temperature
(or entropy) does not play any significant role in determining the
stellar structure. The spherically symmetric solutions of Einstein's
equations for self-gravitating fluids are given by the well-known
Tolman-Opennheimer-Volkoff equations~\cite{TOV}. A generic feature
of these solutions is the existence of a maximum mass for any
equation of state; as the central density is increased beyond the value 
corresponding to the maximum mass, the stars
become unstable towards collapse to a black hole. One criterion
for the stability of a sequence of configurations is the requirement
that the derivative $dM/d\rho_c$ should be positive (the mass should
be an increasing function of the central density). At the point of
instability the fundamental (pulsation) modes become unstable. If 
stability is regained at higher central densities, the modes by
which the stars become unstable towards the eventual collapse belong
to higher-order harmonics.
\begin{figure}
\begin{center}
\includegraphics[height=8.0cm,width=\linewidth,angle=0]{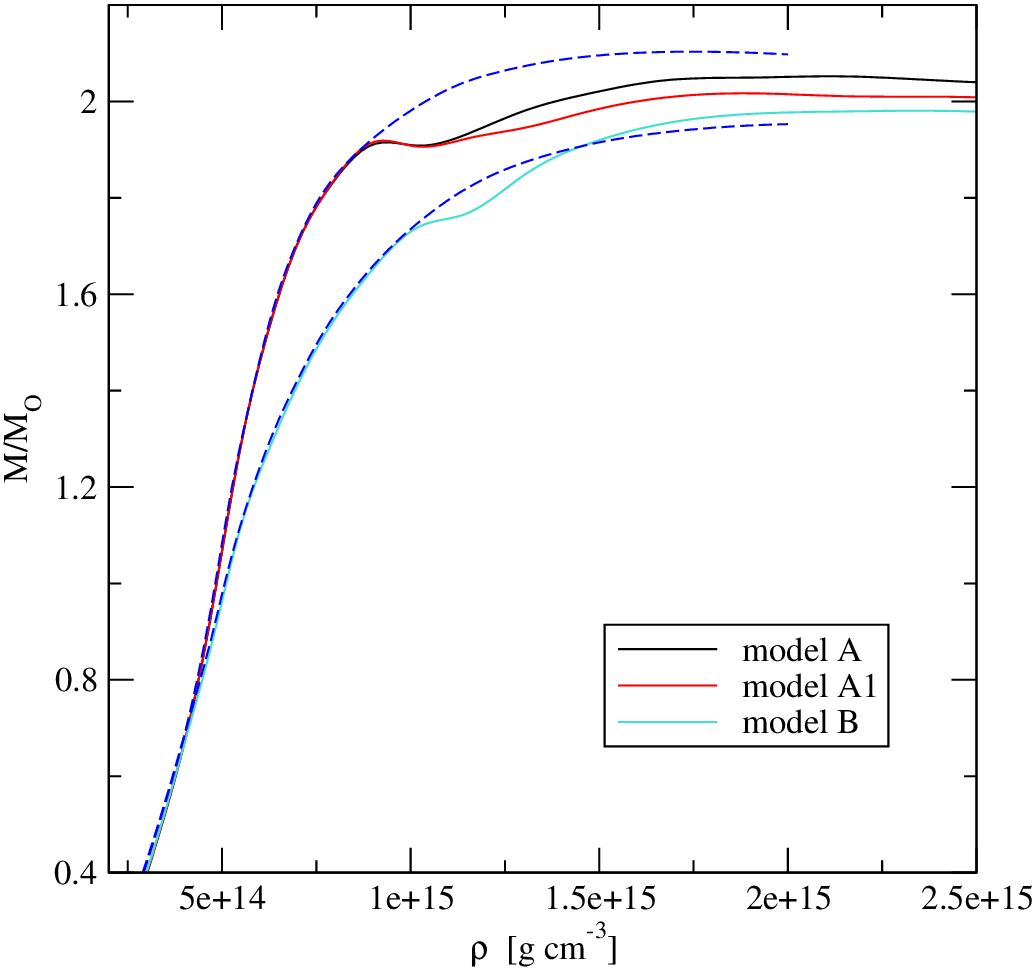}
\end{center}
\caption[]
{Dependence of masses of hybrid, non-rotating compact stars
on their central density for the models A, A1, and B. The dashed lines
show the same for the associated nuclear equations of states.
}
\label{fig:mass_cdens_nonrot}
\end{figure}
For configurations constructed from a purely nuclear equation 
of state the stable sequence extends up to a maximum mass 
of the order 2 $M_{\odot}$ (Fig. \ref{fig:mass_cdens_nonrot}); 
the value of the
maximum mass is large, since our chosen equations of state are
rather hard. The hybrid configurations branch off from the nuclear
configurations when the central density reaches that of the
deconfinement phase transition. The jump in the density at constant
pressure causes a plateau of marginal stability beyond the point
where the hybrid stars bifurcate. This is followed by an unstable
branch ($dM/d\rho_c < 0$). Most importantly, the stability is
regained at larger central densities: a stable branch of hybrid
stars emerges in the range of central densities 
$1.3 \le \rho_c\le 2.5 \times 10^{15}$ g
cm$^{-3}$. The models A and B feature the same high-density quark
matter, whereas the models A and A1 the same nuclear equation of
state. It is seen that the effect of having different nuclear
equations of state (the models A and B) at intermediate densities is
substantial (at densities below $10^{13}$ g cm$^{3}$ all models are
matched to the same equation of state). At the same time, the small
shift $\delta p$ by which the models A and A1 differ does not
influence the masses of stable hybrid stars, although it is
necessary for matching of nuclear and quark EOS in the models A and B.
It is evident that there will exist purely nuclear and hybrid
configurations with different central densities but the same masses.
This is reminiscent of the situation encountered in 
non-superconducting hybrid stars~\cite{Glendenning:1998ag}; the
second branch of hybrid stars was called twins, since for each
hybrid star there always exists a counterpart with the same mass
composed entirely of nuclear matter.

\begin{figure}
\begin{center}
\includegraphics[height=8.0cm,width=\linewidth,angle=0]{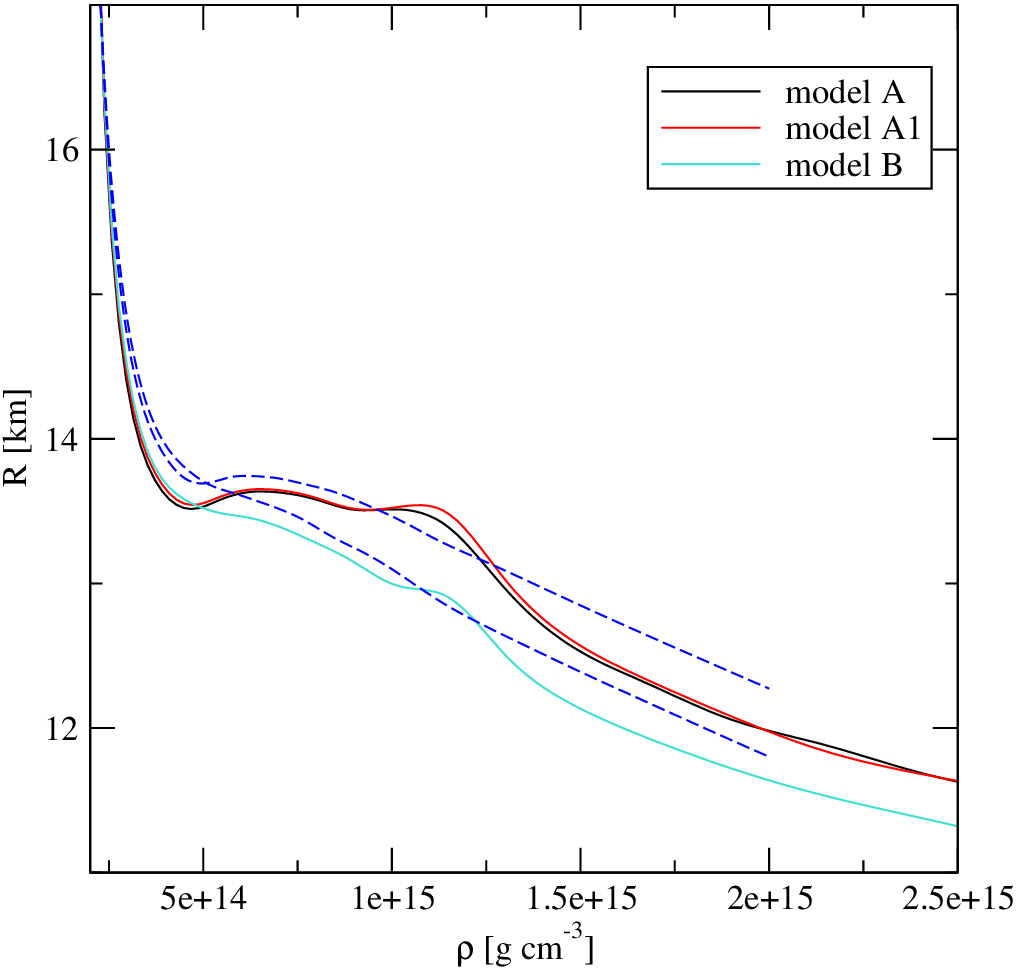}
\end{center}
\caption[]
{Dependence of radii of hybrid, non-rotating compact stars
on their central density for the model A, A1, and B. The dotted lines
show the same for the associated purely nuclear equations of states.
}\label{fig:r_cdens}
\end{figure}
The hybrid configurations are more compact than their nuclear
counterparts, i.~e. they have smaller radii
(Fig.~\ref{fig:r_cdens}). An exception are those configurations 
which are close to the bifurcation point.
However, these belong to the metastable
branch. Contrary to the case of self-bound quark stars, whose radii
could be much smaller than the radii of purely nuclear stars, the
differences between the radii of hybrid  and nuclear stars 
 are insignificant ($\le 1$ km) and cannot be used to distinguish
these two classes by means of current astronomical
observations.

\begin{figure}
\begin{center}
\includegraphics[height=8.0cm,width=\linewidth,angle=0]{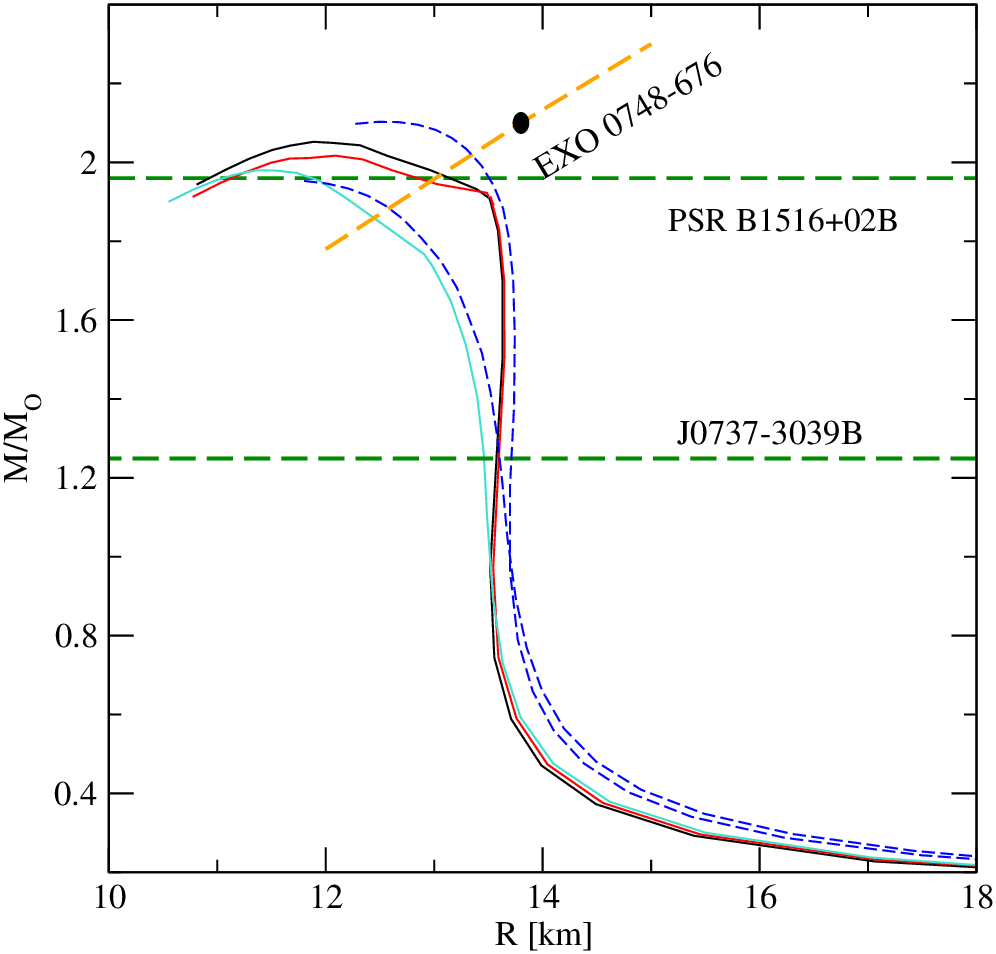}
\end{center}
\caption[]
{
Mass-radius diagram for non-rotating configurations including the bounds 
from EXO 0748-676 and the bounds on the upper and lower pulsar masses.
The sequences of hybrid configurations for model A ({\it heavy, black
  online}), A1 ({\it medium-light, red online}),  and B 
({\it light, blue online}) are shown by solid lines.
Models A and A1 share the same nuclear (low-density) equation of
state, while the models A and B share the same quark (high-density)
equation of state.
The dashed lines are the sequences of purely nuclear stars with  
underlying nuclear equations of state of model A (equivalently A1) 
and model B.
}\label{fig:MR_hybrid}
\end{figure}

Figure \ref{fig:MR_hybrid} displays the astronomical bounds 
on the masses and radii of compact stars along with the ``tracks''
for our models on the mass-radius diagram. All bounds are quoted 
at the 1$\sigma$ level. The bound inferred from EXO0748-676, which 
combines information from redshifted O and F lines, the emitting 
area of X-ray radiation, and the Eddington luminosity, constrain 
the mass and the radius of a compact star to lie on a straight 
line shown in Fig.~\ref{fig:MR_hybrid}~\cite{Ozel:2006bv}. Note that 
multiwavelength observations of EXO0748-676 have led to an alternative 
interpretation of data which suggest a 1.35 $M_{\odot}$ neutron star 
with a main-sequence companion~\cite{PEARSON1,PEARSON2}.
Both the hybrid stars and their nuclear counterparts have their 
masses and radii within these bounds.  Our sequences differ from 
those described in Refs.~\cite{Grigorian:2003vi,Alford:2006vz}. 
Instead of having a narrow region of stable hybrid stars, branching off 
at the bifurcation point from the purely nuclear sequence, we 
obtain configurations which cover a broad range of central densities, 
comparable to those covered by purely nuclear equations of state. 
For sequences constructed from models A and A1  there is a range
of masses and radii that correspond to the new family of stars 
discussed in realtion with Fig.~\ref{fig:mass_cdens_nonrot}. 
The stars belonging to the new family lie to the left from the sharp kink
at the point $ R\simeq 13.5$ km and  $M\simeq 1.9 M_{\odot}$. 
The stable branch of
this new family of stars is separated from the stable nuclear sequence 
by an instability region. The sequences corresponding to model B
do not show such an instability region.  Note that in some 
cases the quark equation of state can be indistinguishable from 
the nuclear equation of state (see, e. g., Ref.~\cite{Alford:2004pf})
in which case quark matter can nucleate at densities that are small
compared to those where the purely nuclear and hybrid configurations 
separate.

Some evidence for massive neutron stars with $M\sim 2M_{\odot}$
has been inferred from astronomical observations.  A massive compact 
star may exist in LMXB 4U 1636-536 with $2.0\pm 0.1 M_{\odot}$
~\cite{barret05}. Recent measurements on PSR B1516+02B in GC M5 gave 
$M  =1.96\pm 0.1 M_{\odot}$~\cite{freire}.
The models A and A1 are consistent with these bounds.  
For the model B these bounds correspond to the stable configuration 
with the largest mass.

For completeness, the lower bound on the neutron star mass
1.249$\pm$0.001 $M_{\odot}$ is shown in Fig.~\ref{fig:MR_hybrid},
which is inferred from the millisecond binary J0737-3039~\cite{Lyne}.
It is seen that the latter bound is relevant only to the low-central-density
stars which are composed of nuclear matter.  It should be noted that
canonical  1.4 $M_{\odot}$ mass pulsars will be purely nuclear if
they are described well by the models A and A1, but will contain quark matter
cores if model B is more appropriate.

\subsection{Rapidly rotating configurations}

\label{sec:RR-confug}
\begin{figure}[t]
\begin{center}
\includegraphics[height=8.0cm,width=\linewidth,angle=0]{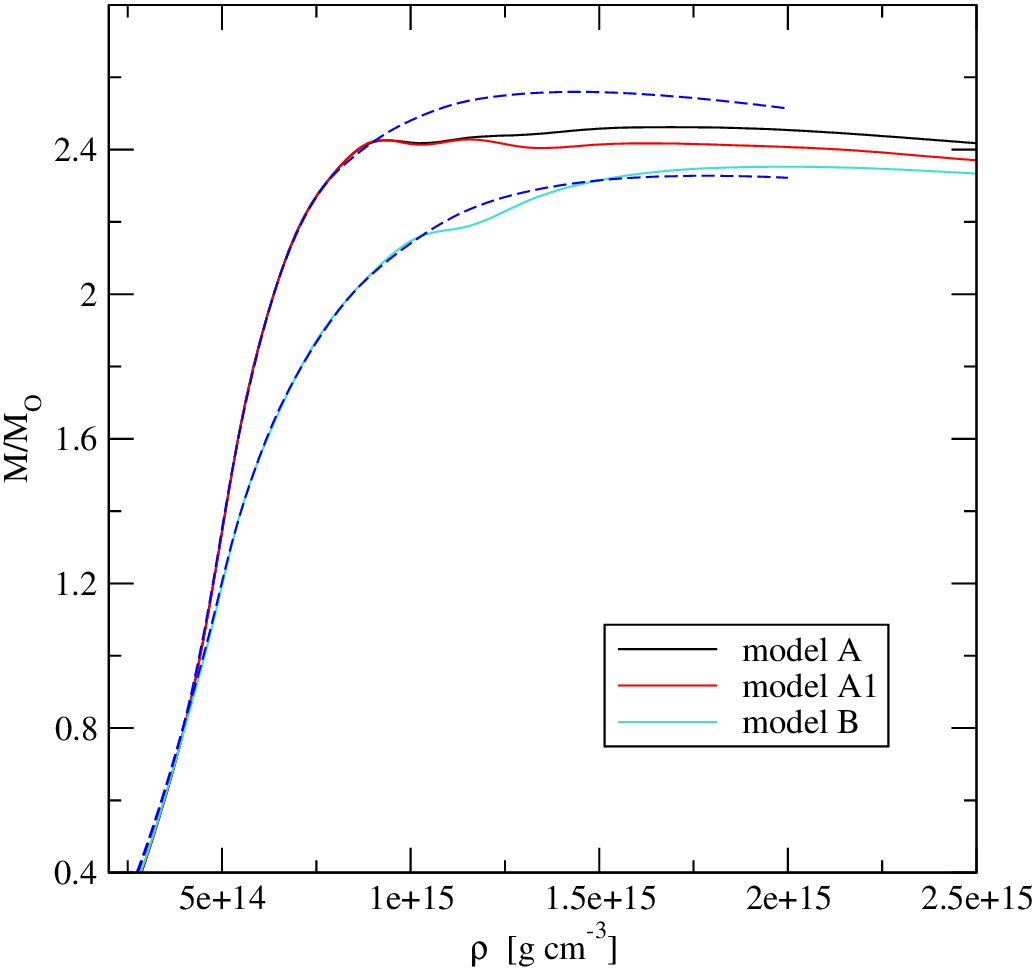}
\end{center}
\caption[]
{Same as in Fig.~\ref{fig:mass_cdens_nonrot} but for stars rotating
at their Keplerian frequency.
}
\label{fig:mass_cdens_kep}
\end{figure}

\vskip 1.5cm

\begin{figure}[t]
\begin{center}
\includegraphics[height=8.0cm,width=\linewidth,angle=0]{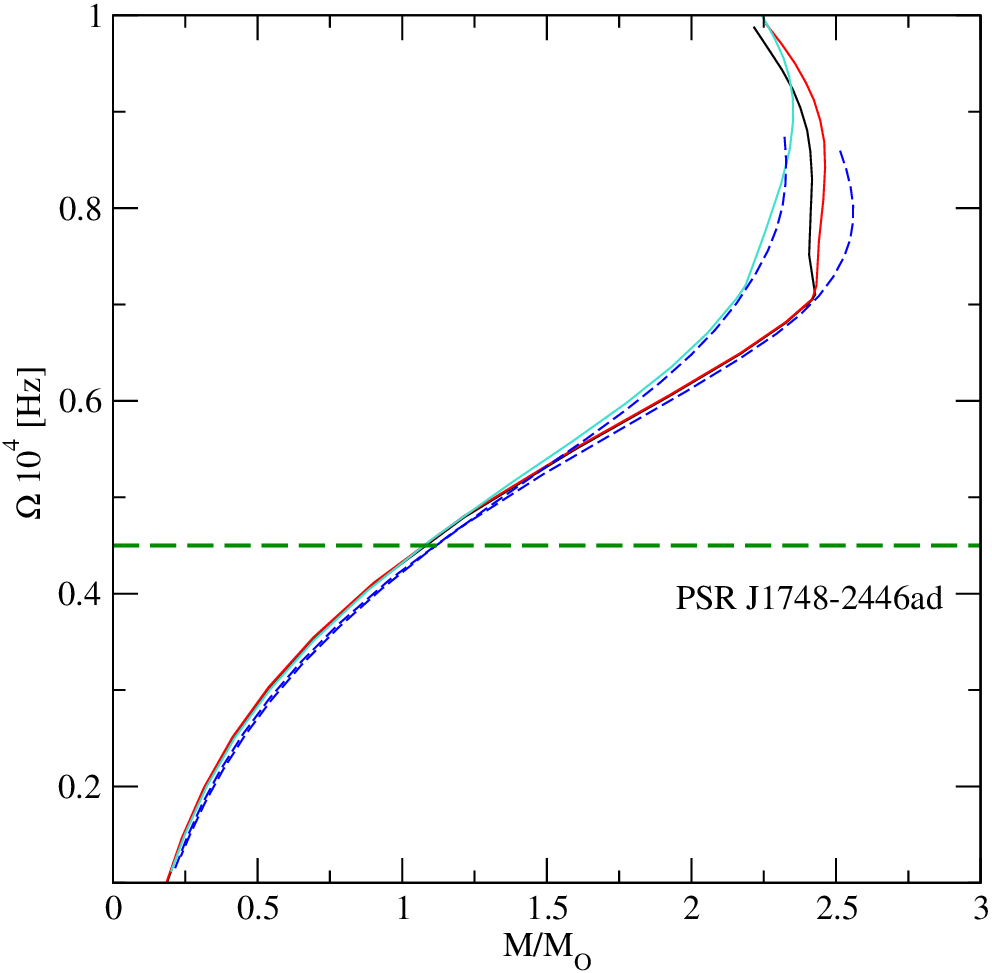}
\end{center}
\caption[]
{Dependence of Keplerian (maximal) frequency of rotation for
hybrid stars on their mass (in solar mass units). The conventions are
the same as in Fig.~\ref{fig:mass_cdens_nonrot}.}
\label{fig:mass_om_kep}
\end{figure}
Millisecond neutron stars can rotate at frequencies which are close to the
limiting orbital Keplerian frequency at which  mass shedding from the
equatorial plane starts. The Keplerian frequency sets an upper limit
on the rotation frequency, since other (less certain) mechanisms, such
as gravitational radiation reaction instabilities, could
impose lower limits for the rotation frequency.
Rapidly rotating pulsars are potentially useful
for placing bounds on the state of  matter at high densities.
Because the centrifugal potential counteracts the compressing stress exerted
on matter by gravity, rotating configurations are more massive than their
non-rotating counterparts. It is assumed that the stars are rotating uniformly:
while nascent hybrid stars can form with large internal circulation,
viscosity and magnetic stress will act to damp such motions, at least
in the non-superfluid component~\cite{Sedrakian:1998ce}. 
The mass versus central-density dependence
of compact stars rotating at the Keplerian frequency is similar to
that for non-rotating stars (compare Figs.~\ref{fig:mass_cdens_nonrot}
and \ref{fig:mass_cdens_kep}) with the scales for mass shifted to
larger values~\cite{RNS}.  The branch corresponding to hybrid
stars is flattened for fast rotating configurations compared to
the non-rotating ones, which implies that small 
variations in the stellar mass can drive  the star unstable.
The increase in the maximum mass for stable
hybrid configurations (in solar mass units) is
$2.052\to 2.462$ for the model A,  $2.017\to 2.428$ and $2.4174$ for
the model A1 (there are two maxima) and  $1.981\to 2.35$ for the model B.

Fast spinning radiopulsars can potentially limit the mass--radius
relation of hybrid stars. The requirement that  the maximum spin frequency
cannot exceed the Keplerian frequency at the neutron star surface
translates into the bound $\nu_{\rm max}\le 1045 (M/M_{\odot})^{1/2} R_{10}^{-3/2}$,
where $R$ is the radius of non-rotating star in units of 10
km~\cite{GRINDLAY,LATTIMER}. The frequency of the fastest radiopulsar
observed to date, PSR J1748-2446ad, is 716 Hz~\cite{Hessels:2006ze}.
The dependence of the orbital Keplerian frequencies of hybrid stars
on their maximum mass is shown in Fig.~\ref{fig:mass_om_kep}. 
The configurations with
masses below 1 $M_{\odot}$ are excluded by the observed frequency 716 Hz.
However, hybrid stars are much more massive - they bifurcate from the
sequence of purely nuclear configurations for $M\ge 2M_{\odot}$. 
Therefore this particular observation is not useful in limiting the 
properties of CCS hybrid stars.

Finally note that at the
point of bifurcation the Keplerian frequency of configurations
jumps to higher values, which is the consequence of the fact that
hybrid stars, having the same mass as their nuclear counterparts, are
more compact and thus can support larger rotation rates.

\begin{figure}
\begin{center}
\includegraphics[height=8.0cm,width=\linewidth,angle=0]{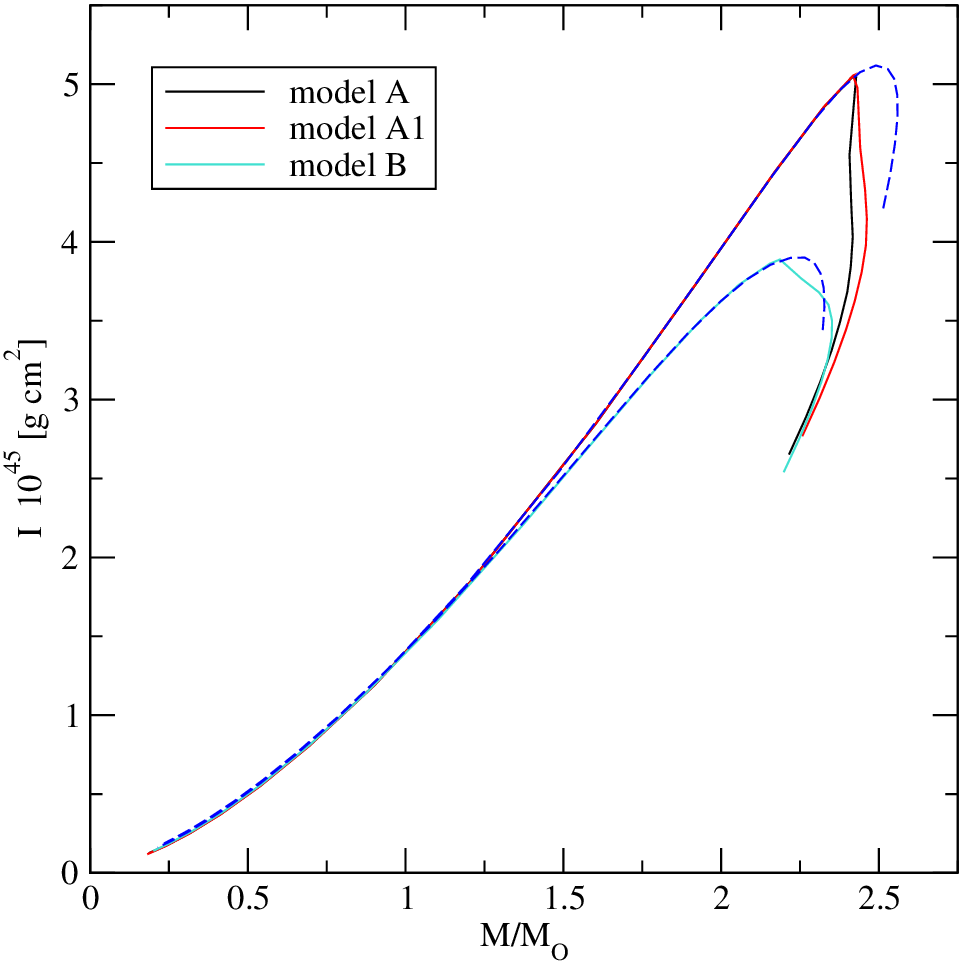}
\end{center}
\caption[]
{Dependence of moment of inertia of the hybrid configurations
on the mass of configuration (in solar mass units)
rotating at the limiting frequency. The conventions are the same
as in Fig.~\ref{fig:mass_cdens_nonrot}.
}\label{fig:mom_in}
\end{figure}

The discovery of the double-pulsar
system PSR J0737-3039 \cite{Lyne,Burgay:2003jj}
offers a unique opportunity to place further bounds on the gross parameters
of compact stars by a measurement of the moment of inertia of star A, since
the spin frequencies and the masses of both pulsars are accurately measured.
Timing measurements over a period of years could provide  information
on spin-orbit coupling which could be revealed through an extra advancement
of the periastron of the orbit above the standard post-Newtonian advance
or in the precession of the orbital plane about the direction of the total
angular momentum~\cite{Lattimer:2004nj}.
The dependence of the moment of inertia, $I$, of configurations
on their mass (in the case of rotation at the Keplerian frequency) is shown
in Fig.~\ref{fig:mom_in}. Since the moment of inertia is independent of the
rotation frequency, we have chosen to extract it for configurations rotating
at the limiting frequency; note that the masses of configurations should be
rescaled appropriately, if one is interested in the $I(M)$ function for slowly
rotating configurations (see ref.~\cite{Morrison:2004df} for a computation of
the moment of inertia in the slow-rotation approximation and a discussion of
the moment of inertia in the context of PSR J0737-3039).
It is seen that while an accurate measurement of the
moment of inertia of pulsar A can
discriminate between the two nuclear equations of state, it
will not be useful for accessing the properties of hybrid stars since the
measured masses of pulsars are too low: $M/M_{\odot} = 1.337$ for pulsar
A and $M/M_{\odot} = 1.250$ for pulsar B; as apparent from
Fig.~\ref{fig:mom_in} observations of heavier objects are
needed to  obtain useful bounds on the moment of inertia of hybrid
configurations. The differences $\sim 20\%$ in the moments
of inertia of purely nuclear and hybrid configurations of the 
same mass
(in the case of the models A and A1) are within the accuracy that can
be achieved in measurements similar to PSR J0737-3039.
Finally we note that the moment of inertia of the CCS phase
is required to compute the strain amplitude of gravitational 
wave emission from an isolated spinning CCS hybrid star. Upper
limits on the strain amplitude have been placed by the LIGO 
and GEO600 detectors~\cite{Abbott:2005,Abbott:2007}. The key 
unknown quantity is the equatorial ellipticity of a hybrid star 
$\epsilon = (I_{xx}-I_{yy})/I_{zz}$, where $I_{ii}$ are the
components of the moment of inertia. The latter quantity has 
been estimated for CCS stars at the qualitative level in 
Ref.~\cite{Lin:2007rz} assuming uniform-density incompressible 
models and maximally strained core~\cite{Ushomirsky:2000ax}. Our 
models of CCS stars can be utilized to obtain quantitative limits
on the strain of gravitational wave emission that can be emitted 
by CCS hybrid stars.

\section{Conclusions}
\label{sec:conclusion}

We constructed a set of equations of states of hybrid stars
which are composed of color-superconducting quark matter at high, and
purely nuclear matter at low densities. High-density quark matter
is described in terms of the semi-microscopic NJL model which includes
pair correlations that lead to the three-flavor LOFF phase as the
ground state of QCD at moderate densities. The low-density nuclear
phase is described in terms of hard relativistic equations of
state based on the Dirac-Bruckner-Hartree-Fock theory.
The sequences of stellar configurations constructed
on the basis of these equations of state
reveal a new branch of {\it stable hybrid configurations} featuring 
CCS matter. These are
analogous to the twin configurations obtained for non-superconducting
quark matter in Refs.~\cite{Glendenning:1998ag} and are in contrast to 
the previously reported NJL-model based sequences of non-rotating 
three-flavor color-superconducting hybrid stars, which were either
entirely unstable or had a narrow range of stable configurations
adjacent to the point where they bifurcate from the
nuclear ones.

A further search in the parameter space is required
in order to judge how generic our constructions are; a quite
general conclusion of our analysis is that the nuclear equations
of states need be {\it hard} to enable the matching to the
quark equations of state. Furthermore, matching is only achieved
if the ratio of the couplings in the diquark and quark-anti-quark channels
is large ($\eta =1$) compared to the value obtained through the Fierz
transformation ($\eta = 0.75$). The lack of confinement in the NJL model
leaves the low-density behavior of the quark matter equation of state
somewhat uncertain. Small variations in the bag constant help in achieving
a phase transition, but do not affect largely the global properties of
hybrid stars (cf. our models A and A1). The maximum masses of our stable
hybrid configurations {\it are consistent with the presently available
astronomical bounds on masses and the
mass-radius relationship of compact stars}.

We provided a first discussion of {\it rapidly rotating} configurations
of CCS
hybrid stars and mapped out their masses and orbital Keplerian frequencies
at which they are destabilized by mass-shedding from the equator. Rapidly
rotating configurations support larger masses (as expected). The mass-central
density relation for this class of objects is ``flatter'' than for their
non-rotating counterparts, i.~e., smaller variations of mass can drive the
system unstable. The Keplerian frequencies of massive members of the
sequences, in particular hybrid configurations, are by a factor of two
larger than the rotation frequency 716 Hz of the fastest millisecond pulsar
known to date. We find that the hybrid configurations, being more compact,
can rotate at substantially larger rates than their nuclear
counterparts of the same mass. The moment of inertia of a hybrid star is
smaller than that of a nuclear star of the same mass. At the point
of bifurcation it drops abruptly giving rise to a sequence of 
stable hybrid stars with almost the same masses but different moments of 
inertia; this implies that simultaneous mass and 
moment-of-inertia measurements 
are useful in distinguishing hybrid configurations from their 
nuclear counterparts.

\section*{Acknowledgments}
We are grateful to R.~Gatto, G.~Nardulli, G.~Pagliara,  
K.~Rajagopal, and the members of the Frankfurt/Darmstadt 
color superconductivity  group for fruitful discussions.
A. S. thanks the Centro di Recerca Matematica 
Ennio De Giorgi, Pisa, Italy, for its hospitality 
during the program ``Many-body theory of inhomogeneous superfluids''.
The research of F.~Weber is supported by the National Science
Foundation under Grant PHY-0457329, and by the Research Corporation.


\end{document}